\documentclass[12pt,preprint]{aastex}
\usepackage{lscape}
\usepackage{longtable}

\begin{document}

\title{Error Analysis of the Light Curve Solution of Contact Binaries Based on the W-D Code}

\author{Liang Liu\altaffilmark{1,2,3,4}}\singlespace

\altaffiltext{1}{Yunnan Observatories, Chinese Academy of Sciences, 396 Yangfangwang, Guandu District, Kunming, 650216, P. R. China (e-mail: LiuL@ynao.ac.cn)}
\altaffiltext{2}{Key Laboratory for the Structure and Evolution of Celestial Objects, Chinese Academy of Sciences, 396 Yangfangwang, Guandu District, Kunming, 650216, P. R. China}
\altaffiltext{3}{Center for Astronomical Mega-Science, Chinese Academy of Sciences, 20A Datun Road, Chaoyang District, Beijing, 100012, P. R. China}
\altaffiltext{4}{University of Chinese Academy of Sciences, Yuquan Road 19\#, Sijingshang Block, 100049 Beijing, China}

\begin{abstract}
We use the idea of repeat measurement to determine the mean value and error of light curve solution parameters of contact binaries. Our simulation is realized by the Monte Carlo algorithm and Wilson-Devinney code. This method can obtain the systematic and random error simultaneously. Within our 48 models, the systematic errors are smaller than the random errors in most case. According to the numerical calculations, it is found that the relative errors of photometric mass ratios are less than 1\,\% for totally eclipsing contact binaries, while they are generally between 10\,\% and 20\,\% for partly eclipsing ones. The effect of third light on the errors of photometric solution is also investigated. With a third light, these errors are close to 10\,\% for totally eclipsing contact binaries. Specially, it is better to set the third light to zero in flux if that light is very faint (e.g., less than 1\,\% contribution in luminosity), because such faint third light will bring big errors to the light curve solutions.
\end{abstract}

\keywords{contact binary stars, mass ratio}

\section{Introduction}
Contact binaries are one type of close binary whose components have overflowed their Roche lobes and shared convective or radiative common envelopes. The W UMa-type contact binaries refer to FGK main-sequence contact binaries which have convective common envelopes \citep[CCEs; e.g.,][]{Lucy1968a,Lucy1968b}. Because of CCEs, thermal timescale mass transfer \citep[TRO,][]{Lucy1976,Flannery1976,RobertsonEggleton1977} and large scale energy transfer \citep[e.g.,][]{Stepien2009} are very important physical processes which seriously affect the evolution of contact binaries. Accurate basic physical parameters (masses, radii and luminosities of each component) are the key information for studying these two physical processes. Furthermore, contact binaries may be progenitors of luminous red novae \citep[LRNs; e.g., V1309 Sco,][]{Tylendaetal2011,Stepien2011,Zhuetal2016,Pietrukowiczetal2017} and blue stragglers \citep[BSs; e.g.,][]{ Eggleton2012,Ferreiraetal2019}. The basic physical parameters of contact binaries are also necessary to understand the formation of these peculiar objects.

To obtain the accurate basic physical parameters of contact binaries, it is necessary to analyze light curves and radial velocity curves simultaneously. We can obtain orbital inclinations via light curves and determine mass ratios as well as mass functions through radial velocity curves. Combining these two results, we can calculate the basic physical parameters. However, due to the circularization and synchronization of orbit, the self-rotational linear velocities of components of a contact binary are almost the same as the orbital velocity. Moreover, because of the short orbital period of contact binaries \citep[a peak of 0.31 days,][]{Qianetal2020}, the rotating velocities of their components are as large as hundreds of kilometers per second, resulting in very serious Doppler broadenings. To make matters worse, such Doppler broadenings are also asymmetric due to geometric asymmetry of components. Thus, the shifts of spectral lines are high probable to be covered by the line broadenings, which limits the accuracy of spectroscopic mass ratio ($q_{\rm{sp}}$). \cite{Rucinski1992,Rucinski2002} had developed the broadening function (BF method) to solve this problem, and the typical relative errors of $q_{\rm{sp}}$ are about 2\,\%.

In addition to the above methods, the basic physical parameters of contact binaries can also be obtained through the combination of light curve and parallax data, but the errors may be larger. This method would work because of the fact that components have filled and overflowed their Roche lobes so that the mass ratio of a contact binary is related to the size of Roche lobe and consequently it could be yielded from light curves. Mass ratios obtained from light curves are called as photometric mass ratios ($q_{\rm{ph}}$). Nevertheless, $q_{\rm{ph}}$ may be reliable only for totally eclipsing contact systems due to the influence of geometric space projection \citep{Pribullaetal2003,TerrellWilson2005}.

Through the above introductions, it can be seen that both of these two methods to determine the basic physical parameters have limitations. We also note that no matter which method is applied, the light curve solution is a necessary process. On the other hand, orbital inclinations and contact degrees only can be obtained by the light curve solution. Hence, it is significance to study how to obtain an accurate solution from the light curve. However, before this paper, except for some individual samples, there was no investigation specifically aimed at analyzing errors of the light curve solution parameters for contact binaries. As mentioned above, due to the limitation, till now the number of the contact binaries which have been obtained $q_{\rm{sp}}$ is less than 200. Even if totally eclipsing contact systems which are potential systems to yield accurate $q_{\rm{ph}}$ are included, the total fraction of such samples still seems less than 20\,\% \citep[according to][]{Lietal2020}. In contrast, \cite{Chenetal2020} have found nearly 350,000 light curves of eclipsing binaries via the latest photometric survey (Zwicky Transient Facility, ZTF for short), including a large number of contact binaries. With the more and more abundant light curves of contact binaries, it is expected that a large number of photometric parameters for contact binaries will be obtained. To evaluate the errors of light curve solutions will become one of the important steps for the further statistical investigation of these photometric parameters. This is also the main goal of this paper.

\section{Method}
We were inspired by the practice of repeat measurement. As we known, the repeat measurement can effectively reduce random errors. The final value of measurement is the mean and the corresponding error is the standard deviation. The process obtaining a light curve for a contact binary is also essentially a measurement process. We can simulate such more complex process numerically.

Any single measurement should include a random error. The measurement of light curve is no exception yet. Let us envisage a situation in which a contact binary system has been observed many times under almost the same observational conditions, and each time its complete light curves have been obtained. It can be predicted that these observed light curves will not be the same exactly because of random errors. Each observed complete light curve will yield a set of independently photometric solution. These solutions have already been affected by random errors. To reduce such errors, we use mean values for each solved photometric parameters. And we take standard deviations as uncertainties for the parameters because the solutions should be Gaussian distributions in an ideal situation. The above process can be achieved with help of the Wilson-Devinney program.

Wilson-Devinney program (W-D for short) is a comprehensive analysis method for solving light and radial velocity curves, which has been developed to the 2015 version \citep{WilsonDevinney1971,Wilson1979,Wilson1990,Wilson2008,Wilson2012,VanHammeWilson2007,Wilsonetal2010,WilsonvanHamme2014}. It includes the LC and DC programs. The LC generates theoretical light or radial velocity curves according to the given parameters, while the DC uses difference correction method to correct the adjusted parameters.

In our method, we use the LC to generate "observed" light curves, and then use the DC to solve them. We generate $N$ sets of light curves, obtaining $N$ groups of solutions. The solutions yield a statistical result. Then, we repeat those steps to another contact binary model. Because the real values of the solution have been already known, we can compute the deviation of the mean value from the true value. Mark the deviation as $\delta_E$, while mark the standard deviation as $\sigma_E$. $\delta_E$ and $\sigma_E$ are corresponding to systematic errors and random errors, respectively, which could be simultaneously determined via our method. Next, we will introduce how to generate ``observed'' light curves and how to solve them.

\subsection{Generating the ``observational'' light curves with the LC}
We perform a simulation for observing a contact binary system of which the period is 0.3 days in filters of $VRI$. The corresponding exposure times are set to 50s for V-band, 30s for R-band and 20s for I-band, respectively. Thus, there are 209 data points for each bandpass. Then, we add a random Gaussian scatter to each generated data point, obtaining a simulated observational multiple colored light curve. We consider the combinations of three typical mass ratios (0.2, 0.4 and 0.6) and inclinations (65$^\circ$, 75$^\circ$ and 85$^\circ$), with two levels of contact degree (20\,\% and 60\,\%). We emphatically focus on the situations that the light curves were affected by a third light. Meanwhile, we take into account several of observational precisions, including different photometric accuracies (0.005, 0.010 and 0.020 mag of the standard deviation) or different time resolutions (209 vs. 500 data points). All given parameters are listed in Table~\ref{tab:given parameters}, where the values of $\Omega_1$ and $L_1/(L_1+L_2)$ are corresponding to different mass ratios or contact degrees.

\subsection{Analyzing the simulated light curves with the DC}
We use the DC to solve the simulated light curves. The adjusted parameters in the DC are: inclination ($i$), mass ratio ($q$), surface potential of star 1 ($\Omega_1$), effective temperature of star 2 ($T_2$), bandpass luminosity of star 1 ($L_{1V}$, $L_{1R}$, and $L_{1I}$), and bandpass flux of third body ($l_{3V}$, $l_{3R}$, and $l_{3I}$) for $l_3 > 0$. If the radiation of third light is isotropic, then we have $L_3 = 4{\pi} \times l_3$ for each single bandpass. The convergence criterion is that all corrections are smaller than the standard deviations and $S_1/|S_1-S_2| < 0.1$, where $S_1$ and $S_2$ are the residual sum of squares calculating from input and predicted parameters, respectively. The input parameters of $q$ and $\Omega_1$ are uniformly randomly assigned within 20\,\% of their true values, while $i$ is set within $\pm7^\circ$ of its true value (if $i > 90^\circ$, then $i = 180^\circ - i$). We also consider a uniform error for $T_1$ (a typical value of $\pm200$\,K) and keep the limb coefficients in line with the temperatures. To save the computing time, if it did not converge when the maximum number of iterations (40) is reached, a new calculation would be started.

It must be pointed out here that we did not attempt to find the best solution for each group of light curves. Once outputs reached the convergence condition, we stopped the calculation and obtained a group of solution. When enough groups of solution (e.g., $N > 3000$) were obtained, distributions of the photometric parameters would not change obviously. Thus, solutions of a certain contact binary model have been completed. Because of the independence of single solution, it is very easy to extend the number of solutions. Although the residual sum of squares did not reach a minimum for a single solution, its distribution always showed a Gaussian profile for any models as we can see in next section.

\section{Results and discussions}
We have made calculations for some typical models of contact binary. The parameters of these models have been listed in Table~\ref{tab:given parameters}. For convenience of description, the short codes being corresponding to the key parameters of each simulated contact binary system are listed in Table~\ref{tab:short code}. Finally, we calculated 48 models of contact binary. The results of solutions are listed in Table~\ref{tab:solution and error} and all corresponding distributions are shown from the appendix Figures A1 to A48 online. The numbers marked in Table~\ref{tab:solution and error} are corresponding to labels of the appendix figures. Each model includes four data rows, which are true values, mean values, standard deviations ($\sigma_E$) and deviations ($\delta_E$, the mean value minus the true value), respectively. $f_S$ and $f_C$ both are contact degrees, but they are yielded from different method. $f_S$ is a statistical mean value of the $N$ numbers of contact degree which is calculated from each single solution ($f=(\Omega_1-\Omega_{\rm{in}})/(\Omega_{\rm{out}}-\Omega_{\rm{in}})$, where $\Omega_{\rm{in}}$ and $\Omega_{\rm{out}}$ are the inner and outer critical potential of Roche lobe, respectively). $f_C$ is a directly computed value which is calculated from the mean value of $\Omega_1$, and the corresponding $\sigma_E$ is computed thorough the error transfer formula. $L_T$ is the total luminosity of $L_1+L_2$ while $L_T^{\prime}$ is the total luminosity of $L_1+L_2+L_3$. From the results shown in Table~\ref{tab:solution and error}, we found that

a) if there is no third light, $f_C$ is usually smaller than $f_S$, but $f_C$ has a very large $\sigma_E$;

b) if the accuracies of data decease, $f_C$ will be larger than $f_S$ (e.g., model\,11 in Table~\ref{tab:solution and error});

c) $\delta_E$ are less than $\sigma_E$ in most cases;

d) $\delta_E$ of the estimated mass ratio display a systematically positive bias in most case.

Remember that in a real solution $\delta_E$ is hard to obtained because the true value of parameter is not sure. Hence we can only use $\sigma_E$ to estimate the relative errors. Fortunately, $\delta_E$ can be ignored in most cases due to the result c). Also note that the residual sum of squares are always good Gaussian distributions, regardless of how irregular distributions of the other parameters (as shown in the online appendix figures), only except one case (model\,23, see section 3.2). This may be a basis for the validity of single solution. In other words, we should reject the solutions of which the residual sums of squares are out off the Gaussian distribution. This will help to improve the accuracy of the final result. An example has been shown in Section 3.2.

We also found a systematical bias of the solution parameters. For example, the estimated mass ratios are systematically larger than their true values, only except for the models\,38 and 39. This bias should not be caused by the adopted initial input values of \emph{q}, because those values are a random uniform distribution and this has been checked. It is not caused by data accuracy either, because it can be found in different data accurate levels. The reason is not clear yet. The models\,38 and\,39 refer to the case that the existed third light was not adjusted in the solution. The presence of third light will reduce the amplitude of light curve. Smaller amplitude can be also yielded from a lower inclination, as well as lower mass ratio. Without adjusting $l_3$, the program is forced to yield lower mass ratios or inclinations for offsetting the decrease in amplitude due to the existed third light. These two models therefore displayed negative $\delta_E$ of mass ratio. The main results and discussions are shown in detail as follows.

\subsection{Without a third light}
\begin{figure}
\begin{center}
	\includegraphics[angle=0,scale=1.6]{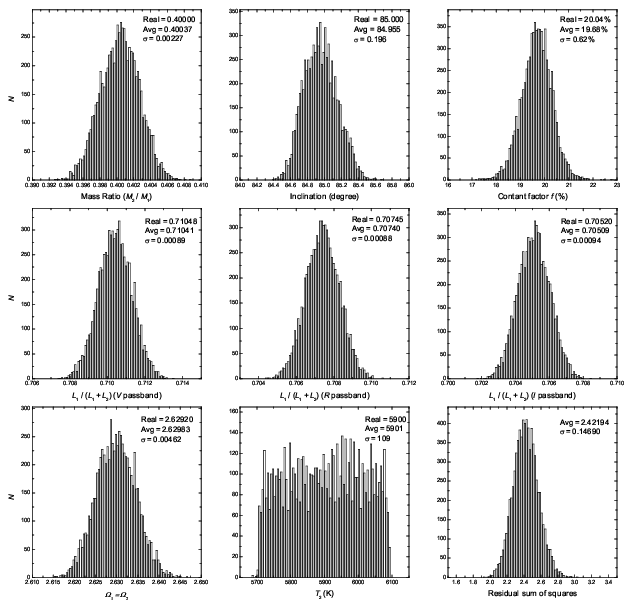}
\end{center}
 \caption{Distribution of the photometric parameters for a totally eclipsing contact binary system with $q=0.4$, $i=85^\circ$, $f=20\%$ and $l_3=0$. The number of the solutions is 7498. NL3, model\,9.}
    \label{fig:nl3d85}
\end{figure}

This is the most common case. Figure~\ref{fig:nl3d85} shows the photometric solution statistics of a contact binary system with a mass ratio of 0.4 and a contact degree of 20\,\%, but without a third light (model\,9 in Table~\ref{tab:solution and error}). In this figure, almost all the parameters are Gaussian distributions. The corresponding mean values and standard deviations ($\sigma_E$) of the parameters are shown in the figure. Readers can calculate $\delta_E$ by themselves or find them in Table~\ref{tab:solution and error}. Through our method, the relative error of the mass ratio of the contact binary is less than 0.6\,\% for this case. It is even smaller (less than 0.1\,\%) if that were calculated with $\delta_E$.

\begin{figure}
\begin{center}
	\includegraphics[angle=0,scale=1.6]{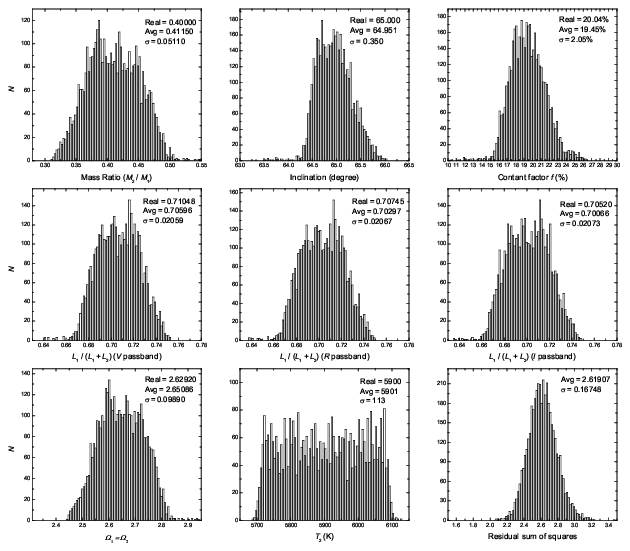}
\end{center}
 \caption{Distribution of the photometric parameters for a partly eclipsing contact binary system with $q=0.4$, $i=65^\circ$, $f=20\%$ and $l_3=0$. The number of the solutions is 4251. NL3, model\,7.}
    \label{fig:nl3d65}
\end{figure}

If the above contact binary is a partially eclipsing system (model\,7 in Table~\ref{tab:solution and error}), the results of the solution are shown in Figure~\ref{fig:nl3d65}. Although the parameters still roughly conform to the Gaussian distribution, they obviously show a larger dispersion than those of the totally eclipsing system. It shows a large $\sigma_E$ and $\delta_E$. The relative error of $q_{\rm{ph}}$ also rises to 12.8\,\% (2.9\,\%, calculated from $\delta_E$).

\begin{figure}
\begin{center}
	\includegraphics[angle=0,scale=1]{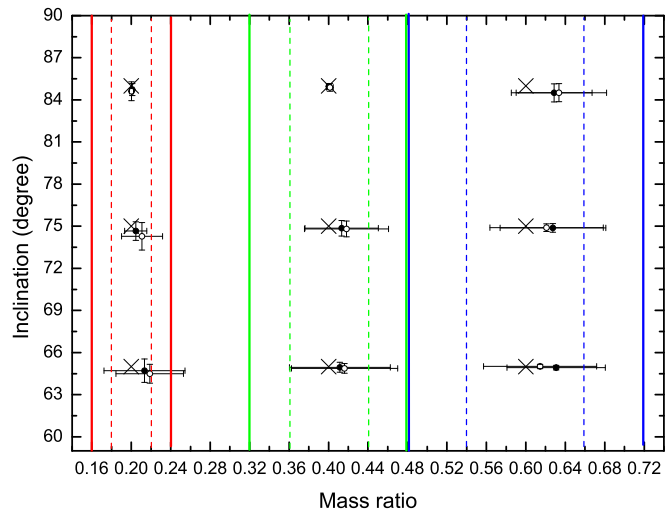}
\end{center}
 \caption{The photometric mass ratios and orbital inclinations for the contact binaries without third lights. The cross symbol refers to the truth value, and the other symbols refer to the calculated values by used of our method. The solid circles denote the contact binary systems with 20\,\% contact degrees, while the empty circles denote those with 60\,\% contact degrees. The vertical solid and dashed lines denote relative errors of 20\,\% and 10\,\%, respectively. Different colors of these lines correspond to different mass ratio truth values.}
    \label{fig:q-i without l3}
\end{figure}

We summarize the solutions of NL3 in Figure~\ref{fig:q-i without l3}. It is found that when $i = 85^\circ$, under the observational accuracy of $\sigma = 0.005$\,mag, the relative error of $q_{\rm{ph}}$ is less than 1\,\%. When $i = 75^\circ$, the relative error of $q_{\rm{ph}}$ is over 10\,\%. Even if the inclination is as low as $65^\circ$, the relative error of $q_{\rm{ph}}$ is still less than 20\,\% in our simulations. The contact degree will affect the relative error of $q_{\rm{ph}}$ slightly.

\subsection{With a third light}
It is also a common phenomenon that a contact binary contains a third body \citep[e.g.,][]{PribullaRucinski2006,D'Angeloetal2006,Rucinskietal2007}, which often contributes more or less to the total flux. Some wide field of view surveys may concentrate several objects into a few pixels, resulting in an image pollution. This situation can also be treated as the case containing a third light.

\begin{figure}
\begin{center}
	\includegraphics[angle=0,scale=1.6]{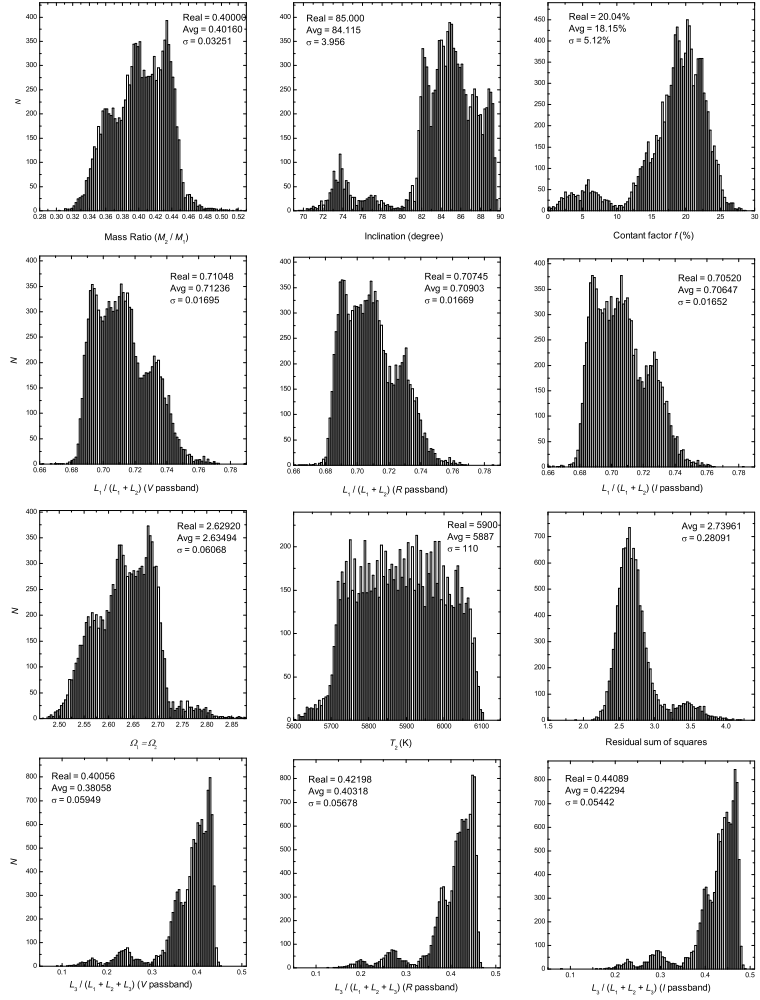}
\end{center}
 \caption{Distribution of the photometric parameters for the totally eclipsing contact binary system with $q=0.4$, $i=85^\circ$, $f=20\%$ and $l_3 \sim 40\,\%$. The number of the solutions is 11535. LL3, model\,23.}
    \label{fig:bl3}
\end{figure}

Figure~\ref{fig:bl3} shows the distributions for one case of LL3 (model\,23 in Table~\ref{tab:solution and error}). In this figure, most distributions have multiple peaks, even the distribution of the residual sum of squares. Fortunately, the primary peaks of these parameters are clear. After having deleted the secondary peak of the residual sum of squares (deleted the values which are greater than 3.3), we indeed obtained an amazing small $\delta_E$ ($-0.00010$ vs. $+0.00106$) for $q_{\rm{ph}}$.

\begin{figure}
\begin{center}
	\includegraphics[angle=0,scale=1]{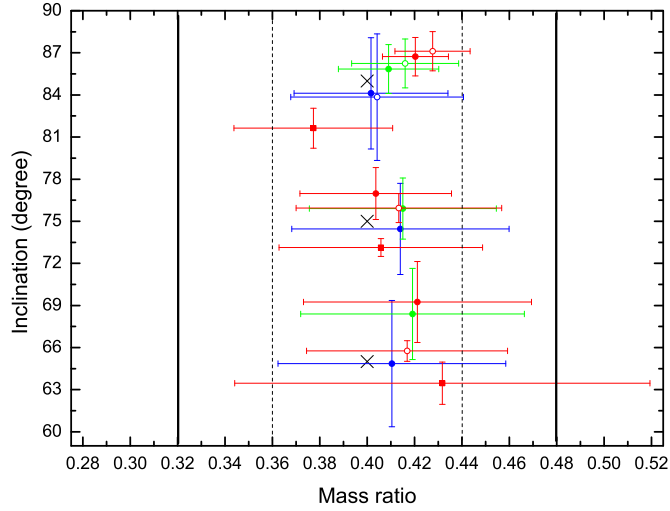}
\end{center}
 \caption{The photometric mass ratios and orbital inclinations for the contact binaries with different third lights. The solid and dotted lines, the cross symbol, the solid circle and empty circle symbols are the same as those in Figure~\ref{fig:q-i without l3}. The blue symbols refer to a large third light (40\,\%, large $l_3$, the case of LL3). The green symbols refer to a small third light (7\,\%, small $l_3$, the case of SL3). Nevertheless all red symbols refer to the cases where the third lights were set incorrectly. The red circle symbols denote that there is a small third light (7\,\%) in the system, but it is not considered in the solutions (fake none small $l_3$, the case of FNSL3). The red square symbols denote that there is no third light (0\,\%) in the system, but it is considered as an adjusted parameter in the solutions (fake $l_3$, the case of FL3).}
    \label{fig:q-i with l3}
\end{figure}

\begin{figure}
\begin{center}
	\includegraphics[angle=0,scale=1]{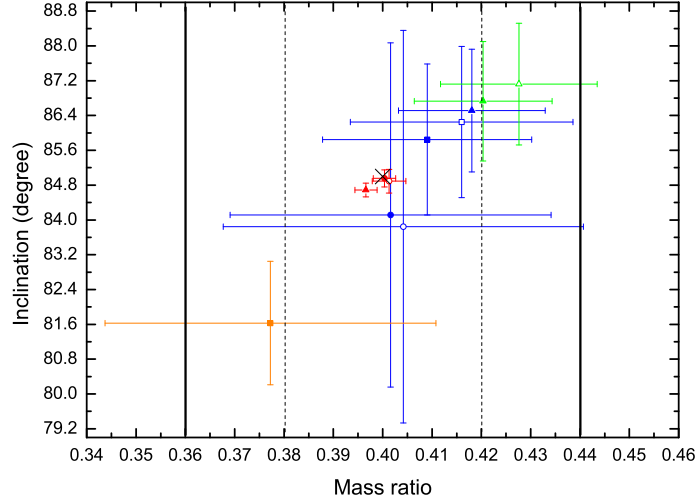}
\end{center}
 \caption{The photometric mass ratios and orbital inclinations for the totally eclipsing contact binary with different third lights. The cross symbol is the same as Figures~\ref{fig:q-i without l3} and~\ref{fig:q-i with l3}. The solid symbols refer to a contact degree of 20\,\%, while the empty symbols refer to a contact degree of 60\,\%, which are also the same as Figures~\ref{fig:q-i without l3} and~\ref{fig:q-i with l3}. The vertical solid lines refer to a 10\,\% relative error, while the vertical dashed lines refer to a 5\,\% relative error. The red symbols denote that none third light was considered in the solutions while the other color symbols indicate completely opposite cases. The red circles refer to the case of NL3, while the red triangles refer to the case of FNVSL3. (There is a very small third light, e.g. 0.6\,\%, in the system, but it is not considered in the solutions.) All blue symbols denote that there is a real third light in the system, and it is also correctly considered in the solutions. The blue circles refer to case of LL3; the blue squares refer to the case of SL3; and the blue triangle refers to the case of VSL3 (0.6\,\%). The green triangles refer to FNSL3. The orange square refers to the case of FL3.}
    \label{fig:q-i of 85 degree}
\end{figure}

Figure~\ref{fig:q-i with l3} and Figure~\ref{fig:q-i of 85 degree} show more cases of contact binaries with third light. It is no doubt that the presence of third light will bring a larger error to the light curve solutions. If there is a third light, $\sigma_E$ increases significantly. On the other hand, for LL3, $\delta_E$ does not change too much (comparing Figure~\ref{fig:q-i without l3} and Figure~\ref{fig:q-i with l3}). This indicates that there is no need to worry too much about the influence of image pollution on the results of the light curve solution. However, with the decreasing proportion of the third light, $\delta_E$ increases very fast (e.g. SL3, the green circles in Figure~\ref{fig:q-i with l3} or the blue squares in Figure~\ref{fig:q-i of 85 degree}). For VSL3, $\delta_E$ is over $\sigma_E$ (the blue solid triangle in Figure~\ref{fig:q-i of 85 degree}). Although the presence of the third light causes a bigger error, the overall relative errors of $q_{\rm{ph}}$ still remain within 15\,\% (Figure~\ref{fig:q-i with l3}).

As mentioned above, the distributions of many solved parameters are no longer Gaussian with the addition of third light. A correlation between mass ratio, inclination and third light may cause this phenomenon. As we known, the amplitude which is a very important characteristic of a light curve mainly depends on those three parameters. The adjusted third light produces many possible solutions of the light curve. Thus, the distribution of inclination splits in several peaks, which is no longer Gaussian. The distribution of mass ratio is also affected. However, we can still take their mean and dispersion as the corresponding value and errors, because these distributions are stable. Through a random sampling test, we found that when $N$ is large enough (e.g. $N > 3000$), the mean and standard deviation are almost constant.

\subsection{Incorrect third light}
Sometimes we are not sure whether the contact system has a third light because the brightness contribution of the third light is too small to distinguish from the data background. It will be true just like an instance that a solar type contact binary contains a third companion of which the spectral type is later than K5. So how do we determine that a contact binary contains such a third light or how can we reduce the interference of the third light in this case? We therefore consider the cases of FL3, FNVSL3, and FNSL3 (Table~\ref{tab:short code}).

\begin{figure}
\begin{center}
	\includegraphics[angle=0,scale=1.6]{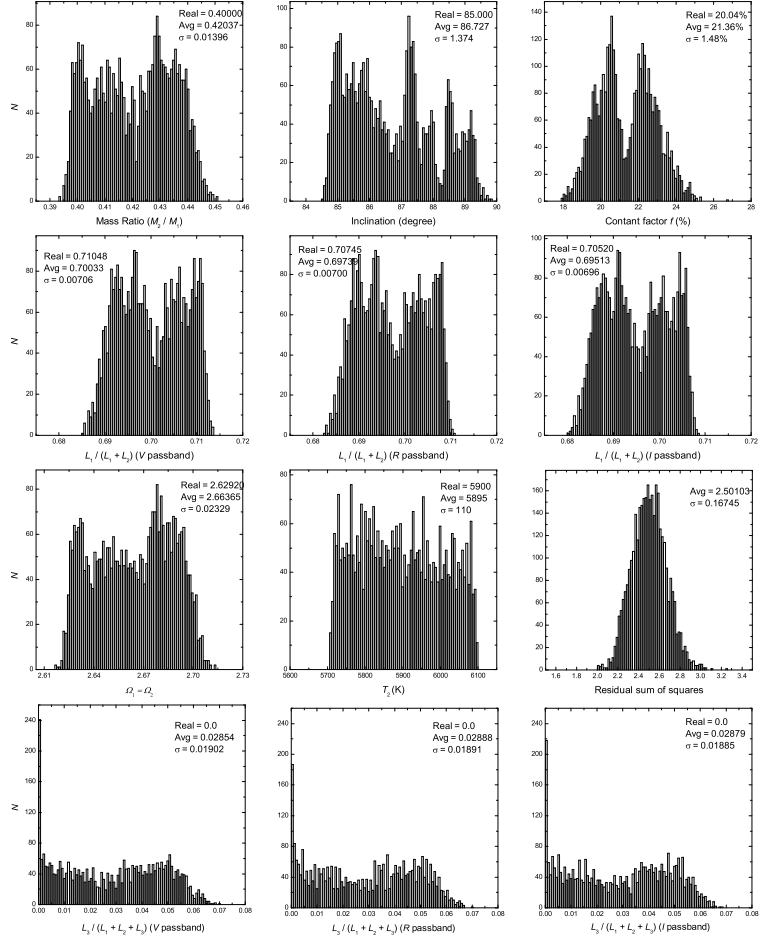}
\end{center}
 \caption{Distribution of the photometric parameters for the totally eclipsing contact binary system with $q=0.4$, $i=85^\circ$, $f=20\%$ and $l_3=0$. However, $l_3$ has been adjusted during the solution. The number of the solutions is 3411. FL3, model\,32.}
    \label{fig:fl3}
\end{figure}

Figure~\ref{fig:fl3} shows the result of an instance of FL3 case (model\,32 in Table~\ref{tab:solution and error}). The photometric parameters deviate from Gaussian distribution seriously, except the residual sum of squares. Compared with the correct treatment of $l_3$ shown in Figure~\ref{fig:nl3d85}, the relative error of $q_{\rm{ph}}$ increased from 0.6\,\% to 3.5\,\%. $\delta_E$ also becomes very large (the red squares in Figure~\ref{fig:q-i with l3}, and the orange square in Figure~\ref{fig:q-i of 85 degree}). If it is FNVSL3 (model\,39 in Table~\ref{tab:solution and error}), the result is the red triangle shown in Figure~\ref{fig:q-i of 85 degree}. However, if we added a correct third light, the result will be the blue triangle of which $\delta_E$ is quite large. It may be due to data noises which have covered such small a third light. For FNSL3, the blue squares denote the correct ones while the green triangles refer to the incorrect ones (Figure~\ref{fig:q-i of 85 degree}). It is obvious that the blue squares have small $\delta_E$. Taking the above results into consideration, to obtain more realistic parameters, we suggest that it should be fixed at zero if the suspected third light is very small (e.g., less than 1\,\%), otherwise it should be set as an adjusted parameter during the solution. This suggestion applies to the totally eclipsing contact binaries. If it is a partially eclipsing system, the situation will become more complex (Figure~\ref{fig:q-i with l3}). The residual of sum squares may be a useful criterion for distinguishing the complicated cases.

\subsection{With different accuracies of the simulated data}
In the actual observations we will obtain data with differently accurate levels. There are two kinds of data accuracy. One is the accuracy of the magnitude and the other is the accuracy of observational time resolution.

\begin{figure}
\begin{center}
	\includegraphics[angle=0,scale=1]{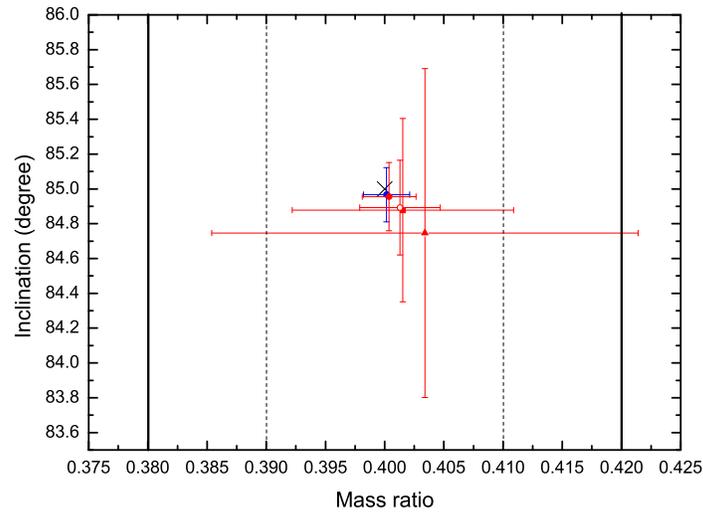}
\end{center}
 \caption{The photometric mass ratios and orbital inclinations for the totally eclipsing contact binary under different data accuracy. The main parameters of the model contact binary system are: $q = 0.4$, $i = 85^\circ$ and $f = 20\,\%$. The red symbols refer to a time resolution of 209 points per cycle, while the blue symbol refers to that of 500 points per cycle (model\,45). The circle, square and triangle symbols refer to accuracies of 0.005, 0.010 and 0.020 magnitudes, respectively. It is showed that both the deviation and standard deviation could be inversely proportional to the magnitude observation accuracy if the time resolutions of the data were the same.}
    \label{fig:q-i with different accuracy}
\end{figure}

Figure~\ref{fig:q-i with different accuracy} reveals the results of simulation solution with the mass ratio of 0.4 and the contact degree of 20\,\% with the different precision of data. The results show that when the time resolutions are the same, higher accuracies of magnitude bring smaller $\delta_E$ and $\sigma_E$. $\delta_E$ and $\sigma_E$ seem to be inversely proportional to the precisions of the magnitude, which is true at least in the given example. It will inspire us to validate more such systems, because such systems could be use to infer the accurate values of the photometric parameters even if the quality of their light curves is low.

\section{Conclusions}
The core idea of this paper is to simulate the process of the repeat measurement in order to reduce the influence of random errors on the light curve solution. We utilized the LC program of W-D code to generate $N$ groups of multiple-colored light curve under a given accuracy. We analyzed these $N$ groups of light curve independently, obtaining $N$ groups of the photometric parameters. Ideally, the distributions of these parameters should be Gaussian profiles. However, in some cases they split into two peaks or multiple peaks. Finally, we adopted the statistical mean values as the measurement values of the photometric parameters, and calculated the corresponding statistical standard deviations. Based on this method, we quantitatively give the errors of the photometric parameters under 48 cases for typical contact binary systems. Main conclusions are as follows:

1) If there is no third light, $q_{\rm{ph}}$ of a totally eclipsing contact binary are highly reliable (the relative errors are better than 1.0\,\%). The relative errors of $q_{\rm{ph}}$ will become from 10\,\% to 20\,\% if the system is a partly eclipsing contact binary.

2) The presence of a third light leads to a large error to $q_{\rm{ph}}$ of a totally eclipsing contact binary. The relative errors become to several percent, but they still less than 10.0\,\%. As for the partly eclipsing contact binary systems, such influence is slight. However, the third light obviously affects the errors of the inclinations.

3) It is not reliable that if the fraction of the third light yielded from the W-D code is very small (e.g., $L_3/L{^\prime}_T < 1\,\%$). In that situation, if the contact binary is also a totally eclipsing system, the relative error of $q_{\rm{ph}}$ could be on the 1\,\% level when the third light was fixed at zero.

4) For a totally eclipsing contact binary system, high time resolution observation does not obviously increase the accuracy of $q_{\rm{ph}}$ when the precision of the magnitude measurement achieve the 0.005 mag. On the contrary, the precisions of the magnitude affect $q_{\rm{ph}}$ very much. Moreover, $\delta_E$ and $\sigma_E$ seem to be inversely proportional to those precisions.

5) The contact degree has an effect on the above results. For the totally eclipsing contact binaries, the bigger contact degrees mean larger errors of the solutions, whether the systems have a third light or not.

\section{Future works}
In this paper, we have discussed some typical cases of contact binaries, including the presence of third light. However, the existence of cool spots is also a very common in contact binaries because of the later spectral type and the rapid rotating. Cool spots models will affect the solutions seriously whereas it will take more computation time to obtain a sufficient result. We also plan to take the phase smearing effect into account. Such effect souring from the long-exposure data is a disaster for analyzing the light curves of contact binaries whose the periods are short. We want to find out the exact errors that will be brought in with this effect. If it were possible, we would apply this method to the semi-detached binaries, even to the detached binaries.

\acknowledgments{We are grateful to the anonymous referee who has given very useful suggestions to improve the paper. We also thank Dr. Nian-Ping Liu for very helpful discussions. This work is partly supported by the National Natural Science Foundation of China (Nos.\,11773066, 11933008), and by the Young Academic and Technology Leaders Project of Yunnan Province (No.\,2015HB098).}


\begin{table}
\begin{center}
\caption{Given parameters of the contact binaries.}
	\label{tab:given parameters}
\begin{tiny}
\begin{tabular}{lcc}
\hline
Parameter                         &  True Value                          &  Offset   \\
\hline
$q=M_2/M_1$                       & 0.2, 0.4, 0.6                        & $\pm 20\,\%$\\
$\Omega_1=\Omega_2$ ($f=20\,\%$)  &  2.2072, 2.6292, 2.9934              & $\pm 20\,\%$\\
$\Omega_1=\Omega_2$ ($f=60\,\%$)  &  2.1562, 2.5316, 2.8528              & $\pm 20\,\%$\\
$i~(^\circ)$                      &  65, 75, 85                          & $\pm 7     $\\
$T_1$ (K)                         &  6000                                & $\pm 200   $\\
$L_1/(L_1+L_2)$ ($V$) ($f=20\,\%$)&  0.81966, 0.71048, 0.63039           & $\pm 0     $\\
$L_1/(L_1+L_2)$ ($R$) ($f=20\,\%$)&  0.81751, 0.70745, 0.62963           & $\pm 0     $\\
$L_1/(L_1+L_2)$ ($I$) ($f=20\,\%$)&  0.81589, 0.70520, 0.62438           & $\pm 0     $\\
$L_1/(L_1+L_2)$ ($V$) ($f=60\,\%$)&  0.80982, 0.69996, 0.62270           & $\pm 0     $\\
$L_1/(L_1+L_2)$ ($R$) ($f=60\,\%$)&  0.80760, 0.69688, 0.61922           & $\pm 0     $\\
$L_1/(L_1+L_2)$ ($I$) ($f=60\,\%$)&  0.80592, 0.69459, 0.61664           & $\pm 0     $\\
$L_3/(L_1+L_2+L_3)$               &  0.0\,\%, 0.6\,\%, 7.0\,\%, 40.0\,\% & $\pm 0     $\\
\hline
\end{tabular}
\end{tiny}
\end{center}
\end{table}

\begin{table}
\begin{center}
\caption{Short codes for the simulated contact binary systems.}
	\label{tab:short code}
\begin{tiny}
\begin{tabular}{ll}
\hline
Short code        &  Condition about the third body ($L_3 = 4{\pi} \times l_3$)\\
\hline
NL3               &  None third body, $l_3$ was fixed to zero and $L_3/L_T {\rm(true)} = 0\,\%$              \\
LL3               &  With a large third body, $l_3$ was adjusted and $L_3/L_T {\rm(true)}\sim40\,\%$       \\
SL3               &  With a small third body, $l_3$ was adjusted and $L_3/L_T {\rm(true)}\sim7\,\%$        \\
VSL3              &  With a very small third body, $l_3$ was adjusted and $L_3/L_T {\rm(true)}\sim0.6\,\%$ \\
FNSL3             &  Fake of none third body, $l_3$ was fixed to zero but $L_3/L_T {\rm(true)}\sim7\,\%$   \\
FNVSL3            &  Fake of none third body, $l_3$ was fixed to zero but $L_3/L_T {\rm(true)}\sim0.6\,\%$ \\
FL3               &  Fake of third body, $l_3$ was adjusted but $L_3/L_T {\rm(true)} = 0\,\%$                \\
\hline
\end{tabular}
\end{tiny}
\end{center}
\end{table}

\begin{landscape}
\begin{table}
\begin{center}
\caption{Statical solution and errors in different cases.}
	\label{tab:solution and error}
\begin{tiny}
\begin{tabular}{ll rrrr rrrr rrrr rr}
\hline
No. &Cases     & $q=\frac{M_2}{M_1}$  & $i$ ($^{\circ}$) & $f_S$ (\%) & $f_C$ (\%) & $T_1$ (K) &$T_2$ (K) & $\Omega_1=\Omega_2$ & $\frac{L_1}{L_T}$ ($V$) & $\frac{L_1}{L_T}$ ($R$) & $\frac{L_1}{L_T}$ ($I$) & $\frac{L_3}{L_T^{\prime}}$ ($V$) & $\frac{L_3}{L_T^{\prime}}$ ($R$) & $\frac{L_3}{L_T^{\prime}}$ ($I$) & $\Sigma(O-C)^2$ \\
\hline
 1&NL3         & $0.20000  $ & $65.000  $ &$20.04 $ & $20.04$  &   $6000 $ & $5900$ & $2.20720   $ & $0.81966  $ & $ 0.81751  $ & $0.81589  $ & $0.00000 $ &  $0.00000 $ &  $0.00000    $ &   $--       $\\
  &$m_E=0.005$ & $0.21342  $ & $64.709  $ &$17.84 $ & $19.82$  &   $5994 $ & $5900$ & $2.23885   $ & $0.81142  $ & $ 0.80938  $ & $0.80780  $ & $0.00000 $ &  $0.00000 $ &  $0.00000    $ &   $2.69001  $\\
  &$DP=209$    & $0.04101  $ & $0.835   $ &$3.51  $ & $66.44$  &   $116  $ & $112 $ & $0.09018   $ & $0.02183  $ & $ 0.02191  $ & $0.02196  $ & $0.00000 $ &  $0.00000 $ &  $0.00000    $ &   $0.29378  $\\
  &$N=3819$    & $0.01342  $ & $-0.291  $ &$-2.16 $ & $-0.22$  &   $-6   $ & $0   $ & $0.03165   $ & $-0.00825 $ & $ -0.00813 $ & $-0.00810 $ & $0.00000 $ &  $0.00000 $ &  $0.00000    $ &   $--       $\\
  &            & $         $ & $        $ &$      $ & $     $  &   $     $ & $    $ & $          $ & $         $ & $          $ & $         $ & $        $ &  $        $ &  $           $ &   $         $\\
 2&            & $0.20000  $ & $75.000  $ &$20.04 $ & $20.04$  &   $6000 $ & $5900$ & $2.20720   $ & $0.81966  $ & $ 0.81751  $ & $0.81589  $ & $0.00000 $ &  $0.00000 $ &  $0.00000    $ &   $--       $\\
  &            & $0.20452  $ & $74.658  $ &$19.05 $ & $20.08$  &   $6000 $ & $5898$ & $2.21772   $ & $0.81701  $ & $ 0.81483  $ & $0.81313  $ & $0.00000 $ &  $0.00000 $ &  $0.00000    $ &   $2.61267  $\\
\end{tabular}
\end{tiny}
\end{center}
\end{table}
\end{landscape}

\begin{landscape}
\addtocounter{table}{-1}
\begin{table}
\caption{$Continued$}
\begin{center}
\begin{tiny}
\begin{tabular}{ll rrrr rrrr rrrr rr}
\hline
No. &Cases     & $q=\frac{M_2}{M_1}$  & $i$ ($^{\circ}$) & $f_S$ (\%) & $f_C$ (\%) & $T_1$ (K) &$T_2$ (K) & $\Omega_1=\Omega_2$ & $\frac{L_1}{L_T}$ ($V$) & $\frac{L_1}{L_T}$ ($R$) & $\frac{L_1}{L_T}$ ($I$) & $\frac{L_3}{L_T^{\prime}}$ ($V$) & $\frac{L_3}{L_T^{\prime}}$ ($R$) & $\frac{L_3}{L_T^{\prime}}$ ($I$) & $\Sigma(O-C)^2$ \\
\hline
12&NL3         & $0.40000  $ & $65.000  $ &$60.04 $ & $60.04$  &   $6000 $ & $5900$ & $2.53160   $ & $0.69996  $ & $ 0.69688  $ & $0.69459  $ & $0.00000 $ &  $0.00000 $ &  $0.00000    $ &   $--       $\\
  &$m_E=0.005$ & $0.41620  $ & $64.880  $ &$58.51 $ & $58.85$  &   $6006 $ & $5904$ & $2.56207   $ & $0.69375  $ & $ 0.69066  $ & $0.68826  $ & $0.00000 $ &  $0.00000 $ &  $0.00000    $ &   $2.61083  $\\
  &$DP=209$    & $0.04181  $ & $0.346   $ &$4.63  $ & $32.25$  &   $115  $ & $116 $ & $0.08160   $ & $0.01510  $ & $ 0.01554  $ & $0.01587  $ & $0.00000 $ &  $0.00000 $ &  $0.00000    $ &   $0.17934  $\\
  &$N=8805$    & $0.01620  $ & $-0.120  $ &$-1.49 $ & $-1.19$  &   $6    $ & $4   $ & $0.03047   $ & $-0.00620 $ & $ -0.00623 $ & $-0.00633 $ & $0.00000 $ &  $0.00000 $ &  $0.00000    $ &   $--       $\\
  &            & $         $ & $        $ &$      $ & $     $  &   $0    $ & $0   $ & $          $ & $         $ & $          $ & $         $ & $        $ &  $        $ &  $           $ &   $         $\\
13&            & $0.40000  $ & $75.000  $ &$60.04 $ & $60.04$  &   $6000 $ & $5900$ & $2.53160   $ & $0.69996  $ & $ 0.69688  $ & $0.69459  $ & $0.00000 $ &  $0.00000 $ &  $0.00000    $ &   $--       $\\
  &            & $0.41811  $ & $74.804  $ &$58.36 $ & $58.77$  &   $6011 $ & $5907$ & $2.56549   $ & $0.69323  $ & $ 0.69008  $ & $0.68764  $ & $0.00000 $ &  $0.00000 $ &  $0.00000    $ &   $2.47722  $\\
\end{tabular}
\end{tiny}
\end{center}
\end{table}
\end{landscape}

\begin{landscape}
\addtocounter{table}{-1}
\begin{table}
\caption{$Continued$}
\begin{center}
\begin{tiny}
\begin{tabular}{ll rrrr rrrr rrrr rr}
\hline
No. &Cases     & $q=\frac{M_2}{M_1}$  & $i$ ($^{\circ}$) & $f_S$ (\%) & $f_C$ (\%) & $T_1$ (K) &$T_2$ (K) & $\Omega_1=\Omega_2$ & $\frac{L_1}{L_T}$ ($V$) & $\frac{L_1}{L_T}$ ($R$) & $\frac{L_1}{L_T}$ ($I$) & $\frac{L_3}{L_T^{\prime}}$ ($V$) & $\frac{L_3}{L_T^{\prime}}$ ($R$) & $\frac{L_3}{L_T^{\prime}}$ ($I$) & $\Sigma(O-C)^2$ \\
\hline
23&LL3         & $0.40000  $ & $  85.000$ &$20.04 $ & $20.04$  &   $6000 $ & $5900$ & $2.62920   $ & $0.71048  $ & $ 0.70745  $ & $0.70520  $ & $0.40056 $ &  $0.42198 $ &  $0.44089    $ &   $--       $\\
  &$m_E=0.005$ & $0.40160  $ & $  84.115$ &$18.15 $ & $18.95$  &   $6000 $ & $5887$ & $2.63494   $ & $0.71236  $ & $ 0.70903  $ & $0.70647  $ & $0.38058 $ &  $0.40318 $ &  $0.42294    $ &   $2.73961  $\\
  &$DP=209$    & $0.03251  $ & $  3.956 $ &$5.12  $ & $24.78$  &   $115  $ & $110 $ & $0.06068   $ & $0.01695  $ & $ 0.01669  $ & $0.01652  $ & $0.05949 $ &  $0.05678 $ &  $0.05442    $ &   $0.30295  $\\
  &$N=11535$   & $0.00160  $ & $  -0.885$ &$-1.85 $ & $-1.09$  &   $0    $ & $-13 $ & $0.00574   $ & $0.00188  $ & $ 0.00158  $ & $0.00127  $ & $-0.01998$ &  $-0.01880$ &  $-0.01795   $ &   $--       $\\
  &            & $         $ & $        $ &$      $ & $     $  &   $     $ & $    $ & $          $ & $         $ & $          $ & $         $ & $        $ &  $        $ &  $           $ &   $         $\\
24&            & $0.40000  $ & $  85.000$ &$60.00 $ & $60.00$  &   $6000 $ & $5900$ & $2.53160   $ & $0.69996  $ & $ 0.69688  $ & $0.69459  $ & $0.39663 $ &  $0.41824 $ &  $0.43714    $ &   $--       $\\
  &            & $0.40419  $ & $  83.847$ &$56.14 $ & $56.99$  &   $6000 $ & $5888$ & $2.54624   $ & $0.70124  $ & $ 0.69790  $ & $0.69531  $ & $0.37169 $ &  $0.39498 $ &  $0.41494    $ &   $2.70150  $\\
\end{tabular}
\end{tiny}
\end{center}
\end{table}
\end{landscape}

\begin{landscape}
\addtocounter{table}{-1}
\begin{table}
\caption{$Continued$}
\begin{center}
\begin{tiny}
\begin{tabular}{ll rrrr rrrr rrrr rr}
\hline
No. &Cases     & $q=\frac{M_2}{M_1}$  & $i$ ($^{\circ}$) & $f_S$ (\%) & $f_C$ (\%) & $T_1$ (K) &$T_2$ (K) & $\Omega_1=\Omega_2$ & $\frac{L_1}{L_T}$ ($V$) & $\frac{L_1}{L_T}$ ($R$) & $\frac{L_1}{L_T}$ ($I$) & $\frac{L_3}{L_T^{\prime}}$ ($V$) & $\frac{L_3}{L_T^{\prime}}$ ($R$) & $\frac{L_3}{L_T^{\prime}}$ ($I$) & $\Sigma(O-C)^2$ \\
\hline
33&FL3         & $0.40000  $ & $65.000  $ &$60.04 $ & $60.04$  &   $6000 $ & $5900$ & $2.53160   $ & $0.69996  $ & $ 0.69688  $ & $0.69459  $ & $0.00000 $ &  $0.00000 $ &  $0.00000    $ &   $--       $\\
  &$m_E=0.005$ & $0.41687  $ & $65.765  $ &$64.59 $ & $65.05$  &   $5984 $ & $5903$ & $2.54750   $ & $0.68831  $ & $ 0.68585  $ & $0.68394  $ & $0.05059 $ &  $0.04998 $ &  $0.04882    $ &   $2.63719  $\\
  &$DP=209$    & $0.04245  $ & $0.734   $ &$5.67  $ & $31.06$  &   $115  $ & $112 $ & $0.07869   $ & $0.01638  $ & $ 0.01672  $ & $0.01698  $ & $0.03931 $ &  $0.03845 $ &  $0.03785    $ &   $0.18003  $\\
  &$N=4601$    & $0.01687  $ & $0.765   $ &$4.59  $ & $5.01 $  &   $-16  $ & $3   $ & $0.01590   $ & $-0.01164 $ & $ -0.01104 $ & $-0.01066 $ & $0.05059 $ &  $0.04998 $ &  $0.04882    $ &   $--       $\\
  &            & $         $ & $        $ &$      $ & $     $  &   $     $ & $    $ & $          $ & $         $ & $          $ & $         $ & $        $ &  $        $ &  $           $ &   $         $\\
34&            & $0.40000  $ & $75.000  $ &$60.04 $ & $60.04$  &   $6000 $ & $5900$ & $2.53160   $ & $0.69996  $ & $ 0.69688  $ & $0.69459  $ & $0.00000 $ &  $0.00000 $ &  $0.00000    $ &   $--       $\\
  &            & $0.41338  $ & $75.935  $ &$63.53 $ & $64.00$  &   $5994 $ & $5906$ & $2.54435   $ & $0.69133  $ & $ 0.68866  $ & $0.68659  $ & $0.03688 $ &  $0.03694 $ &  $0.03615    $ &   $2.53879  $\\
\end{tabular}
\end{tiny}
\end{center}
\end{table}
\end{landscape}

\begin{landscape}
\addtocounter{table}{-1}
\begin{table}
\caption{$Continued$}
\begin{center}
\begin{tiny}
\begin{tabular}{ll rrrr rrrr rrrr rr}
\hline
No. &Cases     & $q=\frac{M_2}{M_1}$  & $i$ ($^{\circ}$) & $f_S$ (\%) & $f_C$ (\%) & $T_1$ (K) &$T_2$ (K) & $\Omega_1=\Omega_2$ & $\frac{L_1}{L_T}$ ($V$) & $\frac{L_1}{L_T}$ ($R$) & $\frac{L_1}{L_T}$ ($I$) & $\frac{L_3}{L_T}$ ($V$) & $\frac{L_3}{L_T}$ ($R$) & $\frac{L_3}{L_T}$ ($I$) & $\Sigma(O-C)^2$ \\
\hline
42&HTS-NL3     & $0.20000  $ & $85.000 $ & $20.04 $ & $20.04 $ & $6000  $ & $5900  $ & $2.20720 $ & $0.81966  $ & $0.81751   $ & $0.81589   $ & $0.00000 $ & $0.00000 $ & $0.00000 $ &        \\
  &$m_E=0.005$ & $0.20082  $ & $84.706 $ & $18.48 $ & $19.39 $ & $5996  $ & $5893  $ & $2.20990 $ & $0.81958  $ & $0.81740   $ & $0.81570   $ & $0.00000 $ & $0.00000 $ & $0.00000 $ &6.06406 \\
  &$DP=500$    & $0.00132  $ & $0.276  $ & $1.37  $ & $3.15  $ & $115   $ & $110   $ & $0.00403 $ & $0.00061  $ & $0.00061   $ & $0.00067   $ & $0.00000 $ & $0.00000 $ & $0.00000 $ &0.24504 \\
  &$N=3415$    & $0.00082  $ & $-0.294 $ & $-1.56 $ & $-0.65 $ & $-4    $ & $-7    $ & $0.00270 $ & $-0.00009 $ & $-0.00011  $ & $-0.00019  $ & $0.00000 $ & $0.00000 $ & $0.00000 $ &--      \\
  &            & $         $ & $       $ & $      $ & $      $ & $      $ & $      $ & $        $ & $         $ & $          $ & $          $ & $        $ & $        $ & $        $ &        \\
43&            & $0.40000  $ & $65.000 $ & $20.04 $ & $20.04 $ & $6000  $ & $5900  $ & $2.62920 $ & $0.71048  $ & $0.70745   $ & $0.70520   $ & $0.00000 $ & $0.00000 $ & $0.00000 $ &        \\
  &            & $0.41128  $ & $64.939 $ & $19.43 $ & $20.06 $ & $6008  $ & $5908  $ & $2.65079 $ & $0.70581  $ & $0.70281   $ & $0.70051   $ & $0.00000 $ & $0.00000 $ & $0.00000 $ &6.23256 \\
\hline
\end{tabular}
\end{tiny}
\end{center}
\end{table}
\end{landscape}

\end{document}